\documentclass{emulateapj}
\usepackage{graphicx}
\usepackage{layout}
\usepackage[latin1]{inputenc}
\usepackage{amssymb}
\usepackage{natbib}

\shorttitle{The low-mass end of the \Mbh{}/\Mhost{} relation in quasars}
\shortauthors{Decarli et al.}

\def\Lsun{L$_\odot$}
\def\Msun{M$_\odot$}
\def\Mbh{$M_{\rm BH}$}
\def\Lhost{$L_{\rm host}$}
\def\Mhost{$M_{\rm host}$}
\def\Mbulge{$M_{\rm bulge}$}

\def\Civ{C\,{\sc iv}}
\def\Mgii{Mg\,{\sc ii}}
\def\Feii{Fe\,{\sc ii}}

\def\kms{km s$^{-1}$}

\def\Hb{H$\beta$}

\def\lsim{\mathrel{\rlap{\lower 3pt \hbox{$\sim$}} \raise 2.0pt \hbox{$<$}}}
\def\gsim{\mathrel{\rlap{\lower 3pt \hbox{$\sim$}} \raise 2.0pt \hbox{$>$}}}

\begin{document}

\title{The low-mass end of the \Mbh{}/\Mhost{} relation in quasars}

\author{
Roberto Decarli\altaffilmark{1}, 
Renato Falomo\altaffilmark{2}, 
Jari K. Kotilainen\altaffilmark{3},
Tomi Hyv\"{o}nen\altaffilmark{3},
Michela Uslenghi\altaffilmark{4}
Aldo Treves\altaffilmark{5}
}
\altaffiltext{1}{Max-Planck Institut f\"{u}r Astronomie, K\"{o}nigstuhl 17, D-69117, Heidelberg, Germany. {\sf decarli@mpia.de}}
\altaffiltext{2}{Osservatorio Astronomico di Padova, INAF, vicolo dell'Osservatorio 5, I-35122 Padova, Italy. {\sf renato.falomo@oapd.inaf.it}}
\altaffiltext{3}{Finnish Centre for Astronomy with ESO (FINCA), University of 
Turku, V\"ais\"al\"antie 20, FI-21500 Piikki\"o, Finland. {\sf jarkot@utu.fi, totahy@utu.fi}}
\altaffiltext{4}{INAF-IASF Milano, Via E. Bassini 15, I-20133 Milano, Italy. {\sf uslenghi@iasf-milano.inaf.it}}
\altaffiltext{5}{Universit\`{a} degli Studi dell'Insubria, via Valleggio 11, I-22100 Como, Italy. Affiliated to INAF and INFN. {\sf aldo.treves@uninsubria.it}}


\begin{abstract}
The \Mbh{}--\Mhost{} relation in quasars has been probed only in a limited
parameter space, namely at \Mbh{}$\sim10^{9}$ \Msun{} and
\Mhost{}$\sim10^{12}$ \Msun{}. Here we present a study of 26 quasars
laying in the low-mass end of the relation, down to \Mbh{}$\sim10^7$ \Msun{}.
We selected quasars from the SDSS and HST-FOS archives, requiring modest 
\Mbh{} (as derived through the virial paradigm). We imaged our sources in 
$H$ band from the Nordic Optical Telescope. The quasar host galaxies have 
been resolved in 25 out of 26 observed targets. Host galaxy luminosities 
and stellar masses are computed, under reasonable assumptions on their star 
formation histories. 
Combining these results with those from our previous studies, we manage to
extend the sampled parameter space of the 
\Mbh{}--\Mhost{} relation in quasars. The relation holds over 2 dex
in both the parameters, similarly to what observed in low-luminosity
AGN and in quiescent galaxies. For the first time, we are able to 
measure the slope of the \Mbh{}--\Mhost{} relation in quasars. We find that
it is consistent with the linear case (similarly to what observed in quiescent 
galaxies).
We do not find any evidence of a population of massive black holes lying 
below the relation.
\end{abstract}

\section{Introduction}

Massive black holes (BHs) are ubiquitously found in the centre of
massive galaxies \citep{kormendy95,decarli07,gallo08}. Their
masses (\Mbh{}) show strong correlations with large-scale properties
of the host galaxies, namely the stellar velocity dispersion 
\citep[$\sigma_*$,][]{gebhardt00,ferrarese00,gultekin09,graham11}, the 
luminosity and mass of the spheroidal component \citep[\Lhost{}, \Mhost{}; see]
[]{magorrian98,marconi03,haering04}. These relations have been interpreted 
as the outcome of a joint evolution between BHs and their host galaxies 
\citep{silk98,kauffmann00,king05,wyithe06,robertson06,hopkins07,malbon07,shankar09}:
In this scenario, the growth of BHs through accretion regulates the gas 
cooling in the outskirt of host galaxies through energy or momentum
injection (feedback), thus quenching the formation of stars.
Galaxy mergers may also play a role in this scenario, as gravitationally-induced
dynamical instabilities may trigger both star formation bursts and gas 
inflows fuelling the BH activity \citep[][although see also Cisternas et al. 
2011]{kauffmann00,dimatteo05,canalizo07,bennert08}.

The \Mbh{}--host galaxy relations have been pinned down on an albeit small
set of local, mostly quiescent galaxies. The sampled parameter space ranges
over 3 dex in masses, from a few million to a few billion solar masses in
terms of \Mbh{}. Extending these studies beyond the local Universe is 
challenging. On one side, the influence radius of BHs, $R_{\rm inf}$, i.e., the
radius where the gravitational potential is dominated by the singularity, is
resolved only in very nearby objects (distances $<$ few tens Mpc) with high 
\Mbh{} values.
For any other sources, indirect tracers of \Mbh{} are required. The most
commonly-adopted indirect estimator of \Mbh{} is
based on the width of broad emission lines and the size of the broad line 
region (BLR). This can be done only in type-1 AGN, where broad lines are observed
\citep{peterson00,vestergaard02,vestergaard06,paperI}. This approach allows
to estimate \Mbh{} from single-epoch spectra in $\sim100\,000$ quasars
up to $z\sim5$ from SDSS spectra \citep{shen10}, and in most of the $z\sim6$
quasars known to day \citep{willott03,kurk07,willott10,derosa11}\footnote{Caveats
to this technique arise as BLR clouds may be supported by radiation 
pressure \citep{marconi09} or move non-virially; projection effects 
depending on the geometry and orientation of the BLR may hinder our ability 
to actually measure the orbital velocity of clouds \citep{nls1,decarli11};
different emission lines may be produced in regions where the gas dynamics 
are different \citep{netzer07,decarli08,labita09,fine11,richards11}.}.

On the other side, the properties of host galaxies are hard to measure in
distant sources. Bright active nuclei (necessary to measure \Mbh{}) can easily 
outshine the light of their host galaxies. Observations in excellent seeing
conditions \citep[e.g.,][]{kotilainen09,targett11} or based on adaptive optics 
\citep[e.g.,][]{croom04,falomo05} are required. Diffraction-limited 
observations with HST have also significantly contributed in this field
\citep{bahcall97,kukula01,ridgway01,hamilton02,dunlop03,floyd04,peng06a,kim08,jahnke09,bennert11,jiang11}, 
although some concerns about the 
reproducibility of the PSF have been arised \citep{kim08}. Up to now, 
$\sim300$ quasar host galaxies have been resolved up to $z\approx3$, and most
of them at $z<0.5$ \citep[see][and references therein]{kotilainen09}.

In order to understand the processes and timescales which
led to the onset of the BH--host galaxy relations, two key 
observational tests are required. The first one consists in 
{\it tracing the evolution of the BH-to-host galaxy mass ratio 
($\Gamma\equiv$ \Mbh{}/\Mhost{}) as a function of Cosmic Time}. If, e.g., 
$\Gamma(z>0)<\Gamma(z\approx0)$, then we can argue that the BHs in these 
systems still have to accrete in already formed bulges. Vice
versa, $\Gamma(z>0)>\Gamma(z\approx0)$ could suggest a rapid growth of the
BHs, followed by a slower build-up of the spheroids.
Most of the studies on the evolution of the BH--host galaxy 
relations suggest that at high redshift, for a given mass of 
the host spheroid, the harbored BH is more massive than at low-$z$ 
\citep{mclure06,peng06a,peng06b,merloni10,paperII,bennert11},
with $\Gamma(z) \propto (1+z)^{3/2}$. It is interesting
to note that the host galaxy of J1148+5251,
the highest-$z$ SDSS quasar, at $z=6.42$, shows a \Mbh{}--host
galaxy dynamical mass ratio of $\approx0.13$ \citep{walter03,walter04},
in an order-of-magnitude agreement with the extrapolation from 
the $z<3$ studies.

A second observational test to probe the onset of the BH--host
galaxy relations is to {\it trace the low-mass end of the
BH--host galaxy relations}. Different initial host galaxy
mass, BH seed mass and build-up processes produce 
different slopes of the relations, especially at low-mass.
Light seeds ($10-100$ \Msun) are expected as the remnants of 
metal-pure stars in the early Universe, while heavier seeds
(up to $\sim 10^5$ \Msun) can result from the direct collapse
of primordial gas clouds. The former ones would produce a larger
scatter in the \Mbh{}--host relations, a higher occupation 
fration in relatively small galaxies and a lower cutoff in the 
minimum \Mbh{} with respect to the latter
\citep[see, e.g.,][and references therein]{volonteri09}. Some authors
even claim that the \Mbh{}--$\sigma_*$ relation itself is just the
upper limit of a broader distribution, with a number of (hard-to-detect)
modest-mass BHs embedded in relatively massive galaxies 
\citep[e.g.][]{batcheldor10}.

The low-mass end of the \Mbh{}--host galaxy relations has been
probed down to few $10^5$ \Mbh{} in quiescent galaxies or in 
low-luminosity AGN at low-$z$ \citep{greene04,greene07,dong07,thorton08,xiao11,jiang11},
suggesting that the \Mbh{}--host relations hold in quiescent
or mildly active galaxies. However, no effort has been attempted
so far to extend this test to higher luminosity AGN.
Quasars are ideal probes of the BH--host galaxy relations at $z>0$. 
However, while $\Gamma$ in quasars
has been measured up to very high redshift, the ranges of \Mbh{}
and \Mhost{} investigated up to date are limited, and comparable 
with the observed scatter of the BH--host
galaxy relations. 
Filling the low-mass end of the \Mbh{}--\Mhost{} relation therefore 
represents a main challenge and a fundamental step in our comprehension
of the BH--host galaxy evolution. In this paper, we present ground-based 
NIR observations of quasars at $z<0.5$ 
selected so that virial \Mbh{} $<10^9$ \Mbh{}. Our imaging campaign 
successfully resolved 25 quasar host galaxies. This enables us to 
directly probe the slope of the \Mbh{}--\Mhost{} relation in quasars.

The structure of this work is the following: in Section 
\ref{sec_sample} we describe the sample. In Section
\ref{sec_spc_analysis} we present the analysis of the spectra
and we derive \Mbh{} in all our sources. The new observations, the 
data reduction and analysis and the results from the imaging campaign
are presented in Section \ref{sec_ima_analysis}.
In Section \ref{sec_discussion} we discuss our results. Conclusions
are summarized in Section \ref{sec_conclusions}. Throughout the paper
we will assume a standard cosmology with $H_0=70$ km s$^{-1}$ Mpc$^{-1}$, 
$\Omega_{\rm m}=0.3$ and $\Omega_{\Lambda}=0.7$.

\section{The sample}\label{sec_sample}
\begin{figure}
\begin{center}
\includegraphics[width=0.5\textwidth]{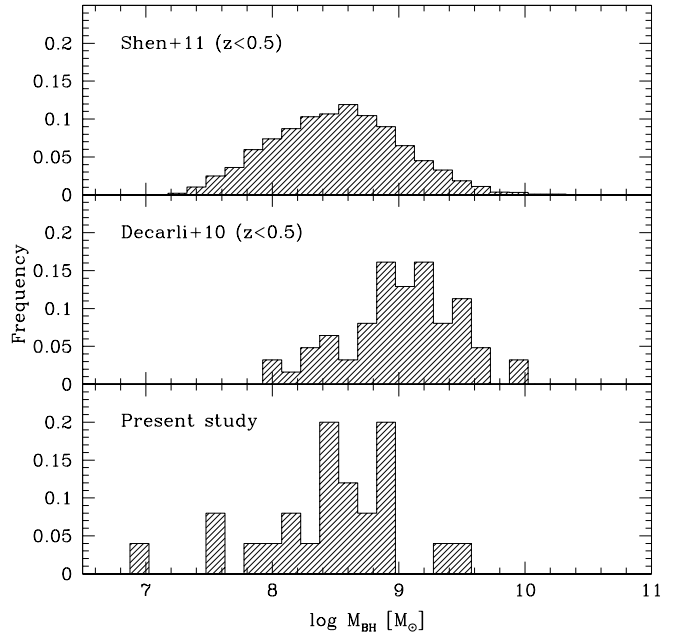}\\
\caption{Comparison among the \Mbh{} distributions in the
SDSS quasars at $z<0.5$ \citep[][; {\it top panel}]{shen10}, in 
our 2010 study ({\it middle panel}) and in the present work
({\it bottom panel}). The latter is clearly more representative
of the general quasar population at low redshift.
}\label{fig_mbh_distr}
\end{center}
\end{figure}

We selected quasars at $z<0.5$ with available \Hb{} observations in
the Sloan Digital Sky Survey \citep[SDSS][]{york00} spectroscopic 
database or \Civ{} or \Mgii{} observations in the HST-Faint Object
Spectrograph (FOS) archive. We require that black hole virial masses, 
computed as described in Section \ref{sec_spc_analysis}, range between 
$10^7$ and $10^9$ \Msun{}. Note that, out of the 62 $z<0.5$ quasars 
examined in our previous study \citep{paperI}, only 8 (13\%) had 
\Mbh{}$<$ $10^9$
\Msun{}. We then selected quasars having at least 3 relatively bright 
($m_H$=10--15 mag) stars within $2'$ (corresponding to the half size 
of the NOTCam field of view), and a number of fainter stars in order 
to have an accurate description of
the Point Spread Function (PSF). Observability constraints and modest
weather losses further limited our analysis to 26 targets (see Table
\ref{tab_sample}), mostly detected in radio wavelengths according
to the \citet{veron10} catalogue. This new sample is then matched
with the 62 $z<0.5$ targets presented in our previous study of the 
\Mbh--\Mhost{} relation \citep{paperI,paperII}. Figure \ref{fig_mbh_distr}
compares the distribution of \Mbh{} in the general SDSS sample at
$z<0.5$, computed from the continuum luminosities and \Hb{} width
estimates reported by \citet{shen10}, with the ones presented in 
\citet{paperI} and in the present study. 

\begin{table}
\caption{{\rm The sample. (1) Target name. (2--3) 
Right ascension and declination (J2000). (4) Is the
target detected in radio wavelengths, according to the
\citet{veron10} catalogue? (5) Catalogue redshift.
(6) Apparent visual magnitude of the quasar.}} \label{tab_sample}   
\begin{center}
\begin{tabular}{ccccccccccc}
   \hline
  Name	            & RA	 &  DEC      &Radio?& $z$	 & $V$   \\
                    &            &           &   &       & [mag] \\
 (1)                & (2)	 &   (3)     &(4)& (5)   & (6)   \\          
 \hline  
PB5723	    & 00 05 47.5 & +02 03 02 & N & 0.234 & 16.60 \\
PG0026+12   & 00 29 13.7 & +13 16 04 & Y & 0.145 & 15.41 \\
PG0052+251  & 00 54 52.1 & +25 25 39 & Y & 0.155 & 15.43 \\
B20110+29   & 01 13 24.2 & +29 58 16 & Y & 0.363 & 17.00 \\
PKS0214+10  & 02 17 07.6 & +11 04 10 & Y & 0.408 & 16.46 \\
J02321+0008 & 02 32 11.8 & +00 08 03 & Y & 0.432 & 19.10 \\
J02331-0909 & 02 33 10.6 & -09 09 40 & Y & 0.388 & 18.45 \\
J03010+0004 & 03 01 00.2 & +00 04 29 & Y & 0.486 & 19.33 \\
J03323+0106 & 03 32 18.0 & +01 06 48 & N & 0.482 & 18.91 \\
J03579-0550 & 03 57 59.0 & -05 50 15 & Y & 0.439 & 18.93 \\
B20752+25A  & 07 55 37.0 & +25 42 39 & Y & 0.446 & 18.00 \\
J08044+1904 & 08 04 42.1 & +19 04 26 & Y & 0.346 & 19.40 \\
J08285+2748 & 08 28 53.5 & +27 48 33 & N & 0.330 & 20.00 \\
J08305+0802 & 08 30 57.4 & +08 02 34 & Y & 0.319 & 19.20 \\
PG0844+349  & 08 47 42.4 & +34 45 03 & Y & 0.064 & 14.50 \\
J09010+3538 & 09 01 00.9 & +35 38 09 & N & 0.302 & 19.20 \\
PG0947+396  & 09 50 48.3 & +39 26 51 & Y & 0.206 & 16.39 \\
PG0953+415  & 09 56 52.4 & +41 15 23 & Y & 0.234 & 15.32 \\
TON1187	    & 10 13 03.1 & +35 51 23 & N & 0.079 & 14.75 \\
TEX1156+213 & 11 59 26.2 & +21 06 56 & Y & 0.349 & 16.90 \\
Q1214+1804  & 12 16 49.0 & +17 48 04 & N & 0.374 & 17.30 \\
PG1404+226  & 14 06 21.9 & +22 23 47 & Y & 0.098 & 15.82 \\
PG1415+451  & 14 17 00.8 & +44 56 06 & Y & 0.114 & 15.24 \\
PG1626+554  & 16 27 56.1 & +55 22 31 & Y & 0.132 & 15.68 \\
4C73.18     & 19 27 48.5 & +73 58 02 & Y & 0.302 & 16.50 \\
PKS2251+11  & 22 54 10.4 & +11 36 39 & Y & 0.325 & 15.82 \\
  \hline
\end{tabular}
   \end{center}
\end{table}

\section{The spectroscopic dataset}\label{sec_spc_analysis}
\begin{figure}
\begin{center}
\includegraphics[width=0.5\textwidth]{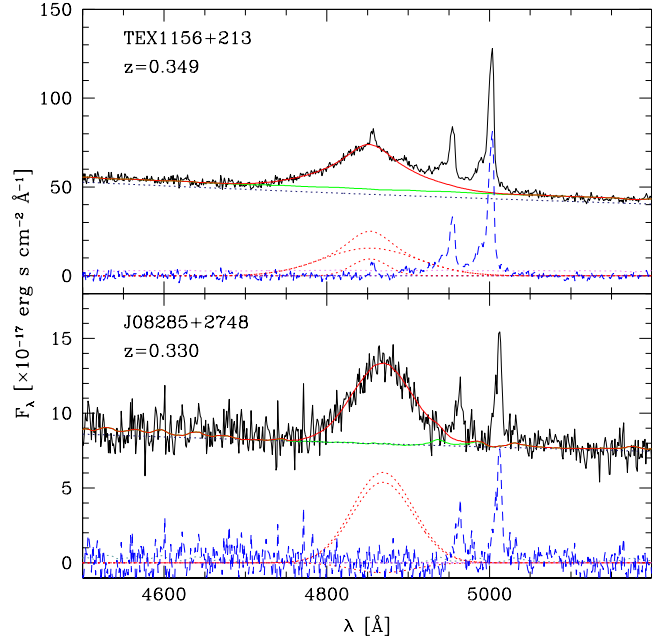}\\
\caption{Examples of H$\beta$ line fitting. The observed spectra, 
shifted to the rest frame, are plotted in solid, black lines.
The various components of the models are shown: the power-law
(black, dotted line); the host galaxy and \Feii{} templates
(green, solid line); the line model (red, solid line) and
its components (red, dotted lines). Fit residuals are shown
in blue, dashed lines.
}\label{fig_spc}
\end{center}
\end{figure}

The spectroscopic dataset consists of pipeline-processed, publicly 
available spectra from the SDSS or FOS archive. SDSS spectra have 
$\lambda/\Delta \lambda \sim 2000$ and a spectral range between 
$3800$ and $9000$ \AA. 
Uncertainties on wavelength calibration amount to $0.05$ \AA, while 
flux calibration formal errors account to 5\%. FOS spectra are
taken from the compilation of re-calibrated quasar and AGN FOS 
spectra by \citet{evans04}. Observations were performed with 
a number of different gratings both at high spectral resolution
(1--6 \AA{} diode$^{-1}$, $\lambda/\Delta \lambda\approx 1300$)
and low spectral resolution (6--25 \AA{} diode$^{-1}$, 
$\lambda/\Delta \lambda\approx 250$) covering various spectral
windows from 1140 \AA{} to 3275 \AA{}. Photometric uncertainties
are usually 5--10\%, while typical wavelength calibration uncertainties
are around 0.5 channels rms \citep[see][]{evans04}.

Spectral analysis follows the same approach described
in \citet{paperI} and \citet{derosa11}. Briefly, continua are modeled
as a superposition of a power-law component, the host galaxy emission
\citep[only in the optical spectra; the Elliptical model by][was adopted
as a template]{mannucci01} and the \Feii{} multiplets \citep[modeled
on the template by][]{vestergaard01}. Relevant broad emission lines
(\Hb{}, \Mgii{} and \Civ{}) are modeled with the superposition of
two gaussian curves with the same peak wavelength. The spectral resolution
of the SDSS data is adequate to allow an easy identification of the
narrow \Hb{} emission with respect to the broad line. Examples
of the line fitting are shown in Figure \ref{fig_spc}.

We measure the continuum luminosity at 1350, 3000 and 5100 \AA{} 
from the fitted power-law. The broad line luminosity and FWHM are
measured from the line model \citep[see][for a discussion on the
fitting technique and the parametrization of the relevant 
quantities]{decarli08}. Virial black hole masses are computed 
with the same recipies used in \citet{paperI}, adopting geometrical
factors of $1.6$ for \Hb{} and \Mgii{} and $2.4$ for \Civ{}.
Table \ref{tab_spc} lists the main measurements and inferred 
quantities from the spectroscopic analysis\footnote{We note that,
after a careful re-analysis of the spectra, two objects (B20110+29
and B20752+25A) show values of \Mbh{} slightly exceeding the 
initial selection criteria. Nevertheless, we include these 
sources in the present analysis.}.
 
\begin{table}
\caption{{\rm Results from the spectroscopic analysis. (1) Quasar name. (2) 
Redshift. (3) Line used in the \Mbh{} estimate. (4) Continuum monochromatic
luminosity at 1350 (for \Civ{}), 3000 (for \Mgii{}) or 5100 \AA{} (for \Hb{}),
in erg s$^{-1}$. (5) Line FWHM in km s$^{-1}$. (6) Virial estimate of the black hole mass, in
solar units.}} \label{tab_spc}   
\begin{center}
\begin{tabular}{ccccccc}
\hline
  Name	            & $z$   & Line & log $\lambda L_\lambda$ & FWHM & log \Mbh{} \\
                    &       &     & [erg s$^{-1}$] & [\kms{}] & [\Msun{}] \\
  (1)               & (2)   & (3) & (4)   &  (5)  & (6)   \\ 
\hline
PB5723	    & 0.234 & \Civ{}  & 44.71 &  3715 & 8.15 \\   
PG0026+12   & 0.145 & \Civ{}  & 45.22 &  2062 & 7.92 \\   
PG0052+251  & 0.155 & \Civ{}  & 45.33 &  5914 & 8.90 \\   
B20110+29   & 0.363 & \Hb{}   & 44.81 &  6149 & 9.33 \\   
PKS0214+10  & 0.408 & \Civ{}  & 45.71 &  4122 & 8.79 \\   
J02321+0008 & 0.432 & \Hb{}   & 44.41 &  1727 & 7.56 \\   
J02331-0909 & 0.388 & \Hb{}   & 44.63 &  1863 & 7.77 \\   
J03010+0004 & 0.486 & \Hb{}   & 44.45 &  6634 & 8.76 \\   
J03323+0106 & 0.482 & \Hb{}   & 44.77 &  4282 & 8.59 \\   
J03579-0550 & 0.439 & \Hb{}   & 44.62 &  4005 & 8.43 \\   
B20752+25A  & 0.446 & \Hb{}   & 45.06 &  9738 & 9.50 \\   
J08044+1904 & 0.346 & \Hb{}   & 44.09 &  7322 & 8.60 \\   
J08285+2748 & 0.330 & \Hb{}   & 44.14 &  5385 & 8.37 \\   
J08305+0802 & 0.319 & \Hb{}   & 44.14 &  6149 & 8.48 \\   
PG0844+349  & 0.064 & \Mgii{} & 44.57 &  3209 & 8.20 \\   
J09010+3538 & 0.302 & \Hb{}   & 44.29 &  8495 & 8.86 \\   
PG0947+396  & 0.206 & \Civ{}  & 45.17 &  4090 & 8.49 \\   
PG0953+415  & 0.234 & \Civ{}  & 45.45 &  3490 & 8.50 \\   
TON1187	    & 0.079 & \Hb{}   & 44.20 &  2141 & 7.60 \\   
TEX1156+213 & 0.349 & \Hb{}   & 44.93 &  5663 & 8.94 \\   
Q1214+1804  & 0.374 & \Hb{}   & 45.04 &  3728 & 8.65 \\   
PG1404+226  & 0.098 & \Hb{}   & 44.14 &  1036 & 6.93 \\   
PG1415+451  & 0.114 & \Hb{}   & 44.19 &  3244 & 7.96 \\   
PG1626+554  & 0.132 & \Civ{}  & 45.18 &  4057 & 8.49 \\   
4C73.18     & 0.302 & \Civ{}  & 45.92 &  4155 & 8.91 \\   
PKS2251+11  & 0.325 & \Civ{}  & 45.54 &  5028 & 8.87 \\   
  \hline
\end{tabular}
   \end{center}
\end{table}

\section{The imaging dataset}\label{sec_ima_analysis}
\begin{figure*}
\begin{center}
\includegraphics[width=\textwidth]{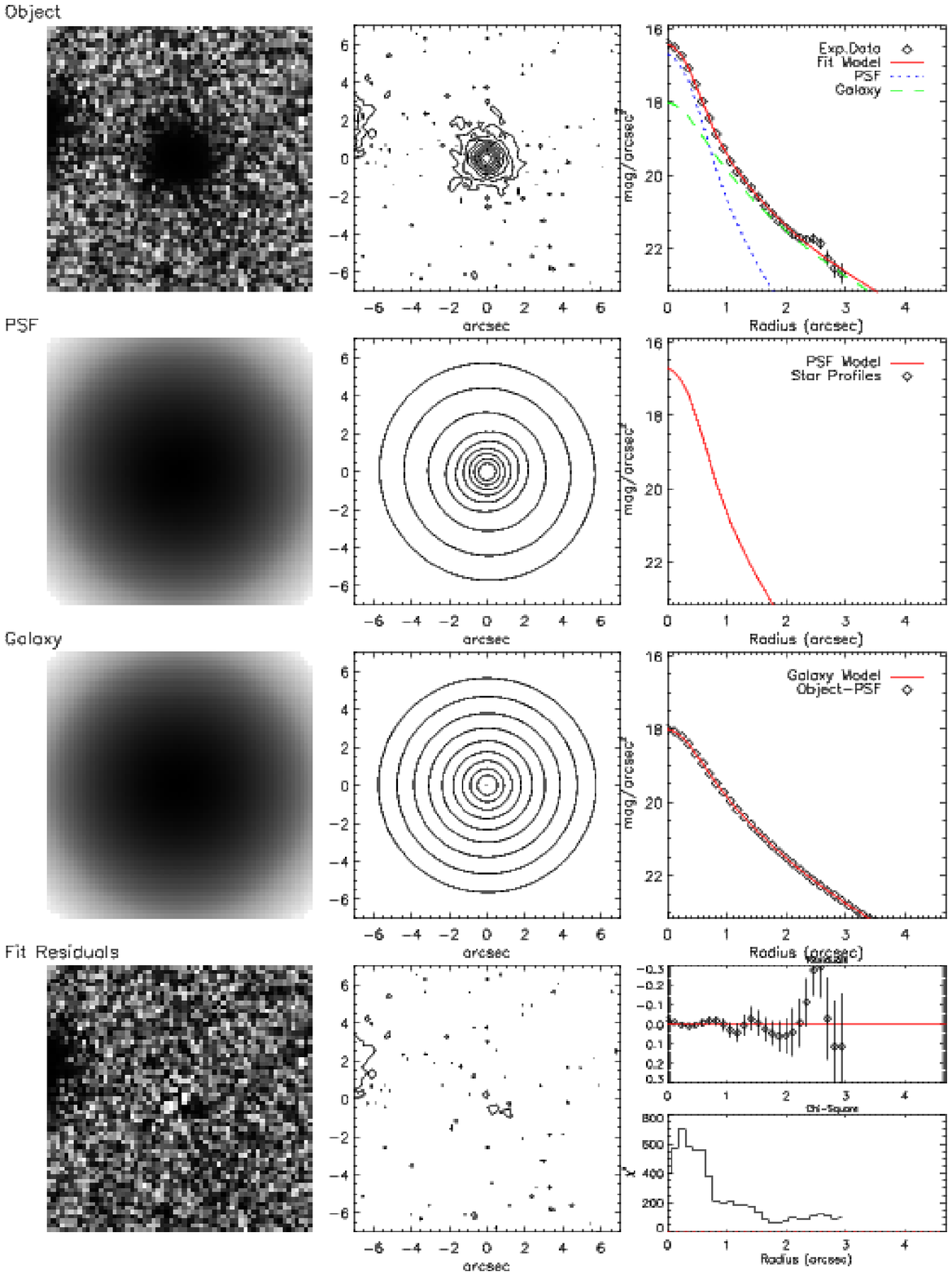}\\
\caption{An example of the analysis of the quasar host galaxies,
shown in the case of the quasar J08285+2748. {\it Top panels:}
Observed quasar image (left), contour plot (center) and light 
profile (right; the PSF and galaxy models are plotted in dotted 
and dashed lines respectively). {\it Central panels:} similarly,
the images (left), contour plots (center) and light profile (right)
of the PSF and the galaxy model. {\it Bottom panels:} image (left),
contour plot (center) and light profile (top right) of residuals
after model subtraction. The radial distribution of $\chi^2$ values 
is also plotted (bottom right).
}\label{fig_ima}
\end{center}
\end{figure*}

All the objects in our study have been observed in $H$ band in a 
campaign at the 2.5m Nordic Optical Telescope (Roque de Los Muchachos, Spain) 
using NOTCam. Observations were carried out during three observing 
runs in May 2007, April and November 2008. The average seeing in $H$
band was $0.7''$. The $1024\times1024$ pixel NOTCam detector has a pixel 
scale of  $0.235''$ pxl$^{-1}$, yielding a field of view size of 
$\sim 4\times4$ arcmin$^2$. 
Usual observing techniques for broad-band NIR imaging of point-like 
sources were adopted. Observations were split in 1 min long individual
frames. Random jittering patterns within a 20$''$ box were adopted in 
order to perform optimal sky subtraction. The total time on each source
was 45 min.

Data reduction was performed using IRAF\footnote{IRAF is distributed by the 
National Optical Astronomy Observatories, which are operated by the 
Association of Universities for Research in Astronomy, Inc., under cooperative 
agreement with the National Science Foundation.}. 
Bad pixels were corrected for in each image using a mask made from the ratio 
of two sky flats with different illumination levels. Sky subtraction was 
performed for each science image using a median averaged frame of all 
the other temporally close frames in a grid of eight exposures. Flat fielding 
was made using normalized median averaged twilight sky frames with different 
illumination levels. Finally, images were aligned to sub-pixel accuracy using 
field stars as reference points and combined after removing spurious 
pixel values to obtain the final reduced co-added image. Zero point 
calibration is achieved by cross-matching the photometry of field stars with 
the 2MASS database. This approach yields typical photometric uncertainties 
$<$0.1 mag.

We analyzed our data using the Astronomical Image Decomposition and 
Analysis package \citep[AIDA;][]{uslenghi08}, an IDL-based
software designed to perform two--dimensional model fitting
of quasar images. Details on the procedure are presented in 
\citet{kotilainen07,kotilainen09}, and briefly summarized here.

A careful modeling of the Point Spread Function (PSF) is
crucial to disentangle the extended host galaxy light from the
nuclear source. To model the PSF shape, we used suitable field 
stars. Each star was modeled with four two-dimensional Gaussians,
representing the core of the PSF, and an exponential feature,
representing the extended wing of the PSF. Regions contaminated by nearby
sources, saturated pixels and other defects affecting the images
were masked out. 

In order to discriminate between resolved and unresolved targets,
we first fit the images of our sources with the pure PSF model.
In most of the cases, an extended halo was clearly observed in
the residuals. We then re-performed the fits using a Sersic law,
(describing the host galaxy) plus a point-source (the nucleus),
convolved to the PSF model. In all but one case (J02331-0909), 
this second fit is significantly better than the fit with the pure 
PSF, as confirmed by the $\chi^2$ ratio between the two fits (see
table \ref{tab_ima}).

An example of the outputs of our analysis is shown in Figure \ref{fig_ima}.

Host galaxy apparent $H$-band magnitudes are then converted into 
rest-frame $R$-band. We use the Elliptical galaxy template by
\citet{mannucci01} to estimate $k$-corrections. The host galaxy
$R$-band absolute magnitude is then converted into a stellar mass
by adopting the mass-to-light ratio ($M/L$) of a single stellar
population originated at $z_{\rm burst}=5$ and passively evolving
down to $z=0$ \citep[see][for details]{paperII}. Table \ref{tab_ima}
summarizes the relevant results from the modeling of the quasar
host galaxies described here.

\begin{table*}
\caption{{\rm Results from the imaging analysis. (1) Quasar name. (2) Redshift.
(3) Apparent observed $H$-band magnitude of the nucleus.
(4) Apparent observed $H$-band magnitude of the host galaxy.
(5) Apparent central surface brightness, as extrapolated from the host galaxy model.
(6) Chi square ratio between the best fit using a pure PSF and 
the best fit using a galaxy+quasar model.
(7) Effective radius of the host galaxy.
(8) Ellipticity of the host galaxy model.
(9) Sersic index of the host galaxy model.
(10) Rest-frame $R$ -- observed $H$ color correction.
(11) Resulting absolute rest-frame $R$-band magnitude of the host galaxy.
(12) Adopted mass-to-light ratio.
(13) Stellar mass of the host galaxy.
(14) BH-to-host galaxy mass ratio.}} \label{tab_ima}
\begin{center}
\begin{tabular}{cccccccccccccc}
\hline
 Name               &  $z$  &  $m_{\rm nuc}$ & $m_{\rm host}$ & $\mu_0$ & $\chi_{\rm psf}/\chi_{\rm gq}$ & $R_e$ & Ell. & $n_{\rm s}$  &  $R$-$H$ & $M_{R}$ & log $M/L$ & log \Mhost{} & log $\Gamma$\\  
                    &       & [mag] & [mag] & [mag $''^{-2}$] &      &      & [$''$] &   &  [mag]   & [mag]   & [\Msun/\Lsun] & [\Msun{}] &  \\   
 (1)                &  (2)  &  (3)  &  (4)  &  (5)  &  (6) &  (7) &  (8) &  (9) &  (10)  & (11  &  (12) & (13)	& (14)      \\
\hline  													     
PB5723	    & 0.234 & 15.99 & 15.81 &  5.41 &  1.41 & 0.32 & 0.57  & 5.00 & 2.59 & -21.93 & 0.71 & 11.38 & -3.61 \\ 
PG0026+12   & 0.145 & 12.90 & 14.94 & 16.10 &  1.77 & 1.44 & 0.15  & 1.20 & 2.58 & -21.64 & 0.75 & 11.30 & -3.76 \\ 
PG0052+251  & 0.155 & 13.46 & 14.37 & 16.34 &  6.55 & 2.19 & 0.19  & 1.22 & 2.58 & -22.38 & 0.74 & 11.59 & -3.07 \\ 
B20110+29   & 0.363 & 16.04 & 16.21 & 12.07 &  2.51 & 3.89 & 0.09  & 5.00 & 2.60 & -22.61 & 0.67 & 11.61 & -2.48 \\ 
PKS0214+10  & 0.408 & 14.60 & 16.10 & 13.98 &  1.88 & 1.16 & 0.00  & 2.77 & 2.59 & -23.03 & 0.65 & 11.77 & -3.36 \\ 
J02321+0008 & 0.432 & 99.90 & 16.80 &  9.93 &  1.43 & 0.37 & 0.00  & 3.89 & 2.58 & -22.50 & 0.64 & 11.54 & -4.18 \\ 
J02331-0909 & 0.388 & 15.98 &  --   &  --   &  1.01 & --   & --    &  --  &  --  &  --    &   -- & --	 &  --   \\ 
J03010+0004 & 0.486 & 99.90 & 17.35 & 13.42 &  1.17 & 0.26 & 0.24  & 1.89 & 2.57 & -22.25 & 0.63 & 11.43 & -2.87 \\ 
J03323+0106 & 0.482 & 18.66 & 17.15 & 11.55 &  1.91 & 0.54 & 0.61  & 3.17 & 2.57 & -22.43 & 0.63 & 11.50 & -3.11 \\ 
J03579-0550 & 0.439 & 18.78 & 17.22 &  8.06 &  1.14 & 0.45 & 0.35  & 5.00 & 2.58 & -22.12 & 0.64 & 11.39 & -3.16 \\ 
B20752+25A  & 0.446 & 14.80 & 16.14 & 12.51 &  1.39 & 0.71 & 0.07  & 2.95 & 2.58 & -23.24 & 0.64 & 11.83 & -2.53 \\ 
J08044+1904 & 0.346 & 17.07 & 17.40 & 17.73 &  1.55 & 0.98 & 0.37  & 1.00 & 2.60 & -21.30 & 0.67 & 11.09 & -2.69 \\ 
J08285+2748 & 0.330 & 16.98 & 17.20 & 16.63 &  1.58 & 0.81 & 0.06  & 1.53 & 2.60 & -21.38 & 0.68 & 11.13 & -2.96 \\ 
J08305+0802 & 0.319 & 17.50 & 18.25 & 17.66 &  1.19 & 0.57 & 0.33  & 0.90 & 2.60 & -20.25 & 0.68 & 10.68 & -2.40 \\ 
PG0844+349  & 0.064 & 13.34 & 14.59 & 17.55 &  1.92 & 2.82 & 0.27  & 0.90 & 2.56 & -20.13 & 0.78 & 10.73 & -2.73 \\ 
J09010+3538 & 0.302 & 17.08 & 16.35 & 14.10 &  2.79 & 0.91 & 0.20  & 2.43 & 2.60 & -22.01 & 0.69 & 11.39 & -2.73 \\ 
PG0947+396  & 0.206 & 14.42 & 15.18 & 16.17 &  4.45 & 1.92 & 0.20  & 1.59 & 2.59 & -22.24 & 0.72 & 11.52 & -3.41 \\ 
PG0953+415  & 0.234 & 12.93 & 15.80 & 19.36 &  1.33 & 3.39 & 0.12  & 0.90 & 2.59 & -21.93 & 0.71 & 11.38 & -3.26 \\ 
TON1187	    & 0.079 & 13.97 & 14.95 & 16.83 &  3.70 & 1.91 & 0.20  & 1.10 & 2.56 & -20.25 & 0.77 & 10.77 & -3.37 \\ 
TEX1156+213 & 0.349 & 15.38 & 15.78 & 14.86 &  2.50 & 1.07 & 0.12  & 1.98 & 2.60 & -22.95 & 0.67 & 11.75 & -3.01 \\ 
Q1214+1804  & 0.374 & 16.36 & 16.48 & 12.05 &  1.11 & 0.25 & 0.40  & 1.97 & 2.59 & -22.44 & 0.66 & 11.54 & -3.09 \\ 
PG1404+226  & 0.098 & 14.35 & 14.70 & 17.08 &  7.62 & 2.99 & 0.52  & 1.04 & 2.57 & -20.97 & 0.77 & 11.05 & -4.32 \\ 
PG1415+451  & 0.114 & 13.65 & 14.02 & 15.22 & 10.56 & 1.50 & 0.04  & 1.30 & 2.57 & -22.01 & 0.76 & 11.46 & -3.70 \\ 
PG1626+554  & 0.132 & 13.80 & 14.78 & 16.23 &  8.55 & 2.04 & 0.04  & 1.53 & 2.57 & -21.60 & 0.75 & 11.29 & -3.18 \\ 
4C73.18     & 0.302 & 13.55 & 16.42 & 17.58 &  1.44 & 1.16 & 0.18  & 0.90 & 2.60 & -21.94 & 0.69 & 11.36 & -2.83 \\ 
PKS2251+11  & 0.325 & 13.53 & 15.70 & 18.62 &  2.94 & 2.43 & 0.05  & 0.90 & 2.60 & -22.86 & 0.68 & 11.72 & -3.23 \\ 
  \hline				        								    
\end{tabular}													   
   \end{center} 												   
\end{table*}

\section{Discussion}\label{sec_discussion}
 
\begin{figure*}
\begin{center}
\includegraphics[width=\textwidth]{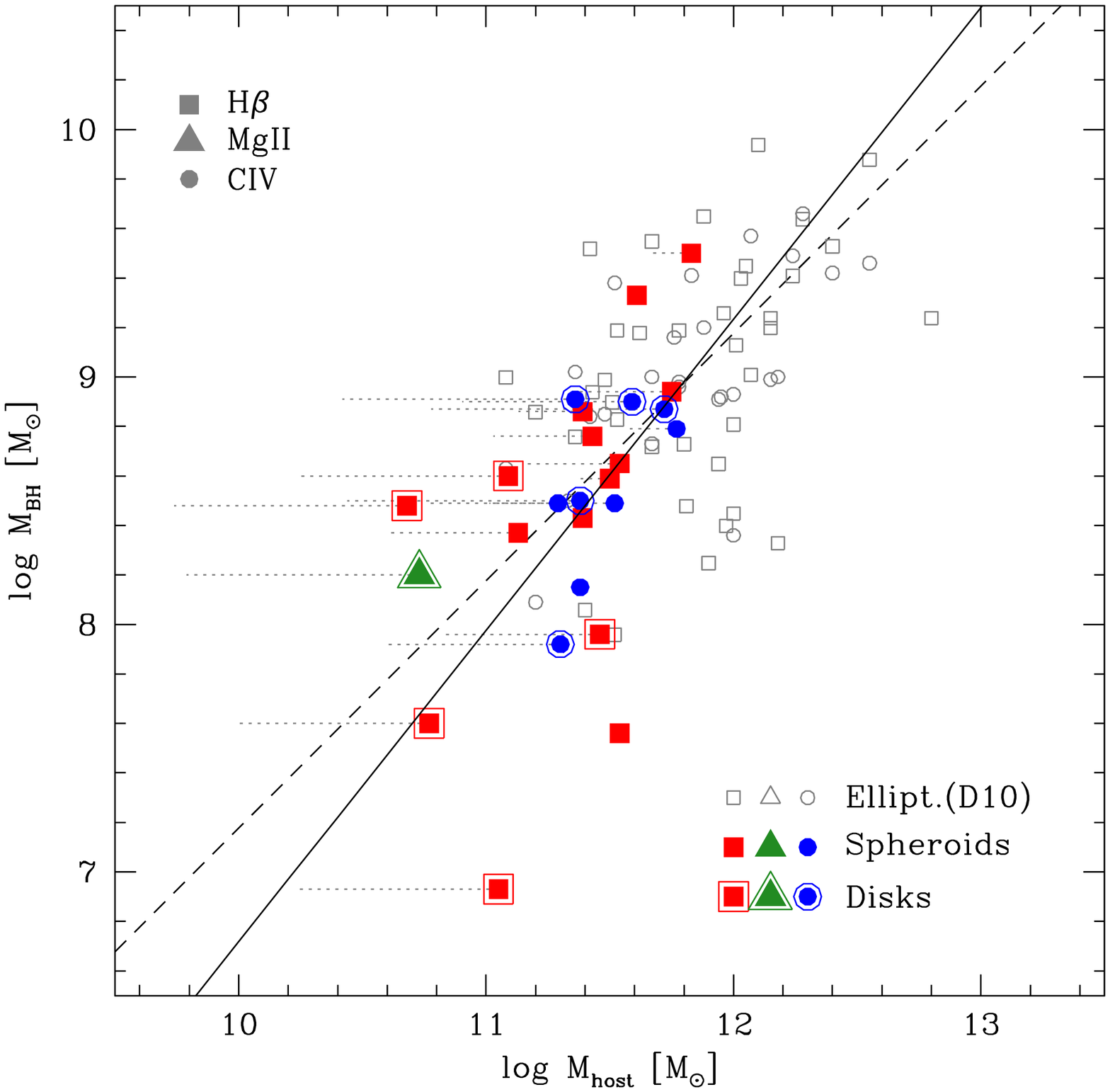}\\
\caption{The \Mbh{}--\Mhost{} relation in the objects in our
sample (filled symbols), as compared with the $z<0.5$ sample
from \citet{paperII} (empty symbols) and the local relation
\citep[\Mbh{}/\Mhost{}=0.0015; see][]{marconi03}.
Squares, triangles and circles refer to \Mbh{} estimates
derived from \Hb{}, \Mgii{} and \Civ{} respectively.
Our new data substantially extend the sampled range
of the \Mbh{}--\Mhost{} relation. The relation holds
down to quasar host masses of $\sim10^{11}$ \Msun{} (BH
masses of $\sim10^8$ \Msun{}). There is no evidence of a 
population of objects lying {\it below} the relation, as
claimed by \citet{batcheldor10}. The overall scatter
of the \Mbh--\Mhost{} relation is 0.44 dex, and it
extends over 2 dex both in terms of host galaxy and
BH mass. Dotted, horizontal lines show how our objects would
move if the correction for the bulge-to-total luminosity
ratio is taken into account (see the text for details).
The solid line is the best fit to our data.}\label{fig_m_m}
\end{center}
\end{figure*}

In Figure \ref{fig_m_m} we show the \Mbh{}--\Mhost{} relation for quasar
host galaxies at $z<0.5$. The dataset (62 objects from Decarli et al. 2010,
plus 25 objects with resolved host galaxies from the present study)
span over 2 dex both in \Mbh{} and \Mhost{}. The same \Mbh{}--\Mhost{} relation
observed for inactive galaxies appears to hold through all the sampled range,
from $\sim 3 \times 10^10$ to $\sim 3\times 10^{12}$ \Msun{} in terms
of \Mhost{}.
We find that $\langle \log \Gamma\rangle=-2.843$ \citep[in excellent 
agreement with the \Mbh{}=0.0015 \Mhost{} value reported by][for inactive 
galaxies]{marconi03} 
with a 0.44 dex scatter. Only 3 sources (J02321+0008, PG1404+226 from this
study; 1001+291 from the old sample) lie more than 2-$\sigma$ {\it below}
the relation. Since the sampled parameter space is about 5 times 
larger than the dispersion of the relation, we can exclude that the observed 
\Mbh{}--\Mhost{} relation is the upper envelope of a
population of quasars with relatively small black holes hosted by 
very massive galaxies.
The best bilinear regression fit of the relation is:
\begin{equation}\label{eq_fit}
\log \frac{M_{\rm BH}}{10^9\, {\rm M_\odot}} = (1.26 \pm 0.29) \times \log \frac{M_{\rm host}}{7 \cdot 10^{11}\, {\rm M_\odot}} + (0.04 \pm 0.03)
\end{equation}
consistent with the relations with a constant \Mbh{}/\Mhost{}
ratio, as observed in quiescent galaxies in the local Universe \citep{marconi03,haering04}.

When considering subsets of our data, the high-mass end shows
a slightly smaller scatter (see Table \ref{tab_gamma}).

For twelve objects the host galaxies are found to be best described by a 
Sersic law with small index ($n_{\rm s}<1.5$), suggesting the presence of significant
disc components. In particular, PG1404+226 (incidentally, the object showing 
the smallest \Mbh{} and the largest deviation with respect to the \Mbh{}--\Mhost{}
relation in our sample) shows clear spiral arms in the residuals of the
host galaxy model. From local galaxy studies, \Mbh{} is found to be
more sensitive to the properties of the spheroidal stellar component rather than 
of the whole galaxy. On the other hand, a bulge+disc decomposition is 
practically impossible with ground-based images of quasar host galaxies
at relatively high redshift.
Here we attempt a rule-of-thumb correction starting from 
the Sersic index value. We assume that the bulge-to-total luminosity ratio in 
the rest-frame $R$ band, $B/T$, scales with the Sersic index as follows:
\begin{equation}\label{eq_bt}
B/T=\left\{\begin{array}{cl}
(n_{\rm s}-0.5)/3.5 & {\rm if}\, n_{\rm s}<4\\ 
1 & {\rm if}\, n_{\rm s}\geq 4
\end{array}\right.
\end{equation}
This simple analytical form roughly traces the bulk of the $B/T$ values for
$n_{\rm s}<4$, as found by \citet{simard11}, who performed
accurate image deconvolution for $\sim 1$ million galaxies from the SDSS.
Furthermore, it is consistent with the operative hypothesis that all the galaxies
well described by a de Vaucouleurs profile ($n_{\rm s}=4$) are bulge 
dominated ($B/T\approx1$), as assumed in our previous study.
The effect of this correction is to move disc-dominated host
galaxies towards the left side of Figure \ref{fig_m_m}. In particular,
all but one source at $M_{\rm bulge}<10^{10}$ \Msun{} would lie 
{\it above} the local relation. The best fit relation is indeed flatter
($0.88 \pm 0.18$ instead of $1.26 \pm 0.29$),
but still consistent with the linear case.
The scatter is also increased (0.53 dex, computed over the whole sample;
0.61 and 0.55 dex in the small \Mbh{} and small \Mbulge{} subsets respectively). 
We stress that the correction reported in equation \ref{eq_bt} is  
uncertain, because of the wide range of $B/T$ values reported for any given 
$n_{\rm s}$. However, we remark that any correction for the $B/T$ would 
make the case against a population of under-massive black holes in very 
massive galaxies even stronger.

A similar argument can be used to evaluate how our results are affected
by different assumptions on the star formation history. In our study, 
we adopted the mass-to-light ratio ($M/L$) of a single stellar
population originated at $z_{\rm burst}=5$ and passively evolving
down to $z=0$. However, objects with significant disc contaminations
are expected to have a younger stellar population than old, passive 
spheroids. This would imply smaller $M/L$, i.e., less massive host galaxies
for a given observed host luminosity. This would make the case
against a quasar population lying {\it below} the observed \Mbh--\Mhost{}
relation even more robust.

\begin{table}
\caption{{\rm Average values of the \Mbh{}/\Mhost{} ratio
in various subsets of our sample. (1) Subsample. (2) Number of
objects. (3) Average value of log $\Gamma$. (4) Root Mean Square
of log $\Gamma$.}} \label{tab_gamma}
\begin{center}
\begin{tabular}{cccc}
\hline
 Subsample          &   N   & $\langle \log \Gamma \rangle$ & RMS \\  
                    &       &  dex  &  dex    \\   
 (1)                &  (2)  &  (3)  &  (4)    \\
\hline  													     
\multicolumn{4}{l}{\it Without $B/T$ correction:}                   \\
ALL  	    	                       & 87 & -2.842 & 0.444  \\
\Mbh{} $\geq 10^9$ \Msun{}             & 33 & -2.627 & 0.351  \\
\Mbh{} $< 10^9$ \Msun{}                & 54 & -2.974 & 0.445  \\
\Mhost{} $\geq 4\times10^{11}$ \Msun{} & 49 & -2.889 & 0.401  \\
\Mhost{} $< 4\times10^{11}$ \Msun{}    & 38 & -2.782 & 0.492  \\
\hline				    
\multicolumn{4}{l}{\it With $B/T$ correction:}                   \\
ALL				       & 87 & -2.697 & 0.524  \\ 
\Mbh{} $\geq 10^9$ \Msun{}	       & 33 & -2.623 & 0.356  \\ 
\Mbh{} $< 10^9$ \Msun{} 	       & 54 & -2.886 & 0.603  \\ 
\Mbulge{} $\geq 4\times10^{11}$ \Msun{} & 46 & -2.886 & 0.419  \\ 
\Mbulge{} $< 4\times10^{11}$ \Msun{}    & 41 & -2.485 & 0.554  \\ 
  \hline													     
\end{tabular}													      
   \end{center}
\end{table}

\section{Conclusions}\label{sec_conclusions}

We measured black hole masses and host galaxy luminosities in 
a sample of 25 low-redshift ($z<0.5$) quasars selected to have
modest ($<10^9$ \Msun{}) black hole masses. For each object we inferred 
stellar masses. This allowed us to significantly expand the sampled 
range of \Mhost{} and \Mbh{} for quasars. We found that:
\begin{itemize}
\item[{\it i-}] The \Mbh{}--\Mhost{} relation holds over all the
2 dex both in terms of \Mbh{} and \Mhost{}. The relation has a 
scatter of 0.44 dex, i.e., the sampled parameter space is $\sim$5 
times larger.
\item[{\it ii-}] The slope of the \Mbh{}--\Mhost{} relation in quasars
is consistent with unity (in a log-log plane), consistently with
what observed in quiescent galaxies.
\item[{\it iii-}] The scatter of the relation increases by $\sim0.9$
dex at the low-mass end.
\item[{\it iv-}] After applying a simplistic correction for the disc 
contribution in objects with low Sersic indexes, the slope of the
\Mbh{}--\Mhost{} relation is smaller but still consistent with the
linear case.
\item[{\it v-}] No evidence of a population of quasars
with relatively modest \Mbh{} and very high \Mhost{} values is found.
\end{itemize}
Further studies at even lower \Mbh{} masses ($\lsim 10^7$ \Msun) and
at higher redshift could provide further constraints on the
early black hole growth and the nature of the seeds, and to pin
down the evolution of the \Mbh{}--host galaxy relations along the
Cosmic Time. This requires extremely high quality imaging of quasar 
host galaxies that would become possible with the next generation of 
ELT and laser assisted AO imagers.

\section*{Acknowledgments}

RD acknowledges funding from Germany's national research center for 
aeronautics and space (DLR, project FKZ 50 OR 1104).
Based on observations made with the Nordic Optical Telescope, operated
on the island of La Palma jointly by Denmark, Finland, Iceland,
Norway, and Sweden, in the Spanish Observatorio del Roque de los
Muchachos of the Instituto de Astrofisica de Canarias. 
This research has made use of the NASA/IPAC Extragalactic Database (NED)
which is operated by the Jet Propulsion Laboratory, California
Institute of Technology, under contract with the National Aeronautics
and Space Administration. 

Facilities:
\facility{NOT(NOTCAM)}
\facility{SDSS}
\facility{HST}

\label{lastpage}

\end{document}